\documentclass[prb]{revtex4}

\usepackage{graphicx}
\usepackage{dcolumn}
\usepackage{bm}

\begin{document}

\preprint{APS/123-QED}

\title{Layer interdependence of transport in an undoped Electron-Hole Bilayer}

\author{Christian P. Morath}\email{cpmorat@sandia.gov}

\author{John A. Seamons}%

\author{John L. Reno}%

\author{Mike P. Lilly}

\affiliation{
Sandia National Laboratories, Albuquerque, NM 87185
}%

\email{cpmorat@sandia.gov}
\altaffiliation[Also at ]{Physics and Astronomy Department, University of New Mexico.}

\date{\today}

\begin{abstract}
The layer interdependence of transport in an undoped electron-hole bilayer (uEHBL) device was studied as a function of carrier density, interlayer electric field, and temperature.  The uEHBL device consisted of a density tunable, independently contacted two-dimensional electron gas (2DEG) and two-dimensional hole gas (2DHG) induced via field effect in distinct GaAs quantum wells separated by a 30 nm Al$_{0.9}$Ga$_{0.1}$As barrier.  Transport measurements were made simultaneously on each layer using the van der Pauw method.  An increase in 2DHG mobility with increasing 2DEG density was observed, while the 2DEG mobility showed negligible dependence on the 2DHG density.  Decreasing the interlayer electric-field and thereby increasing interlayer separation also increased the 2DHG mobility with negligible effects on the 2DEG mobility.  The change in interlayer separation as interlayer electric-field changed was estimated using 2DHG Coulomb drag measurements.  The results were consistent with mobility of each layer being only indirectly dependent on the adjacent layer density and dominated by background impurity scattering.  Temperature dependencies were also determined for the resistivity of each layer.
\end{abstract}


\maketitle

\section{Introduction}
\label{Section1}

Interest in electron-hole bilayers necessarily arose from the prospect of observing Bose-Einstein condenstion (BEC) of excitons in semiconductor double quantum well systems  \cite{Shevchenko1976, Lozovik1976} and significant progress towards this goal has been made.\cite{Butov2004, Seamons2008}  This trend in bilayer research centered on the behavior of the electron-hole pair.  The work presented in this paper, however, focuses on the individual transport in each layer and the layer interdependence.  The latter is the primary question we seek to answer here and for which a bilayer device is singularly, exceptionally suited; to what extent will the mere presence of a nearby 2DEG affect the transport in a 2DHG and vice-versa?


The general transport properties of the 2DEG (or 2DHG) system in modulation-doped heterostructures were well established decades ago.\cite{Ando1982,Walukiewicz1984} Exploiting Coulomb scattering's dependence on the shape of the wavefunction, Hirakawa \textit{et al.} and others demonstrated later that the 2DEG's mobility dependence on density could be altered by deforming the wavefunction using external fields from gates.\cite{Hirakawa1985,Kurobe1993b, Kurobe1994}  Calculations by Kurobe showed that the 2DEG wavefunction can be \textit{squeezed} by making the back-gate voltage more negative, which narrows the 2DEG's confining potential and moves the wavefunction towards the interface.\cite{Kurobe1993}  Furthermore, he demonstrated that squeezing the wavefunction  reduced remote and space impurities scattering times, but enhanced the channel impurity scattering time.  To eliminate remote-impurity scattering, which typically limits the mobility in modulation-doped heterojunctions, and more closely inspect the roles of other scattering mechanisms, similar studies were also done using undoped, inverted semiconductor-insulator-semiconductor or ISIS structures, pioneered by Meirav \textit{et al}.\cite{Meirav1988} From these studies it was determined that background channel impurities dominate scattering at low densities, while interface roughness dominates at higher densities.\cite{Markus1994, Pettersen1996} More recently, Das Sarma \textit{et al.} and others showed that background impurity scattering in GaAs heterostructures is rudimentary to the 2D metal-insulator transition, which occurs as density is reduced and screening of the random potential landscape, due to these impurities, becomes progressively weaker.\cite{DasSarma2005, Manfra2007}  These studies were all on unipolar devices and, to the author's knowledge, an experimental study of whether general 2D transport is affected by a 2D system of opposite charge in close proximity has not been reported.

In this paper, the results of an investigation into the layer interdependence of transport in a uEHBL are presented. The uEHBL device under study consists of a density tunable, independently contacted 2DEG and 2DHG induced via field effect in distinct GaAs quantum wells separated by a 30 nm Al$_{0.9}$Ga$_{0.1}$As barrier.\cite{Seamons2007}  To populate the undoped wells an interlayer electric field E$_{IL}$ is necessarily established to account for the energy difference between the conduction and valence bands.  This design affords the following advantages: (1) independent contacts allow for simultaneous transport measurements of each layer and Coulomb drag measurements; (2) a tunable density 2DEG and 2DHG allows for these measurements to be made as functions of the density in each layer, $n$ and $p$; (3) an undoped structure reduces or eliminates scattering by remote ionized impurities; and, finally, (4) for the same densities, the distance between the 2DEG and 2DHG or interlayer separation $d$ can be varied by changing E$_{IL}$ and both gate voltages.

The investigation included mobility and resistivity measurements measured in each layer as functions of $n$ and $p$, E$_{IL}$, and temperature $T$.  The results indicated that 2DHG transport changed by varying $n$ or E$_{IL}$, while the 2DEG was largely immune to similar changes in $p$ or E$_{IL}$.  Coulomb drag measurements were used to estimate the change in $d$ as E$_{IL}$ was varied.  These results demonstrate that transport in an uEHBL has an asymmetric layer interdependence, with only the 2DHG transport having an apparent dependence on the adjacent layer density, $n$.  However, this effect appears to be only indirect since increasing $n$ necessitated an increase in the tilting of the 2DHG confinement potential as well. Presumably, this tilting squeezed the hole wavefunction, and thereby reduced the background impurity scattering.  The temperature sweeps of the resistivity of each layer also reflected qualitatively unique behaviors, with the 2DHG displaying  stronger changes at similar densities then the 2DEG.

\section{Material and Fabrication}
\label{Section2}

\begin{figure}
\begin{center}
\includegraphics[width = 8.5cm]{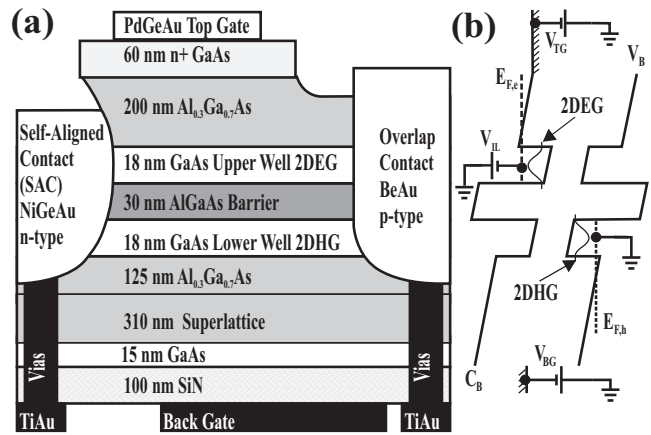}
\end{center}
\caption{\label{fig0} (a) Schematic of the uEHBL device. (b) Cartoon schematic of the energy level diagram showing relative locations of V$_{TG}$, V$_{IL}$, and V$_{BG}$}
\end{figure}

A full account of the design and fabrication of the device used in this study was given in \cite{Seamons2007}.  The uEHBL device was formed from molecular beam epitaxially (MBE) grown GaAs/AlGaAs double quantum well material (wafer EA1286).  A side profile of the device after full processing is shown in Fig. \ref{fig0}a.  The top and bottom 18 nm GaAs quantum wells were separated by a 30 nm Al$_{0.9}$Ga$_{0.1}$As barrier.  Above the top quantum well is a 200 nm Al$_{0.3}$Ga$_{0.7}$As cladding layer and a 60 nm n+ GaAs cap layer.  Beneath the bottom quantum well is a 125 nm Al$_{0.3}$Ga$_{0.7}$As cladding layer and a 310 nm growth superlattice.  Beneath that is a 15 nm GaAs layer, which acts as the second etch stop during backside processing and effectively becomes the cap layer.  As this layer gets exposed to air, the Fermi level at its surface is expected to be pinned at mid-gap.  To process the uEHBL device a $\sim 25$ cm$^{2}$ piece of this material was cleaved from the wafer and mesa-etched in the shape of a Hall bar with 5 arms extending from each side.

The carriers in each quantum well were induced via external fields generated by gates on the top and bottom of the device, as depicted in \ref{fig0}(a).  Each gate covered the central Hall bar region and extended along the entire length of half the arms in a geometry which allows for both R$_{xx}$ and Hall measurements on both wells, independently.  The gate contact pads were positioned at opposite ends of the long axis of the central Hall bar region.

The only intentional dopants present in the material were in the n+ cap layer, which was also used to form the top gate.  These dopants do not populate the upper quantum well.  A shallow annealing of PdGeAu metal was used to make ohmic contact to the top gate.  Directly adjacent to the edge of the top-gate at each arm's end were self-aligned\cite{Kane1993}, n-type NiGeAu ohmic contacts.  These contacts served a dual role as reservoirs supplying electrons to the top quantum well and as the n-type ohmic contacts for transport measurements on the 2DEG.  At the ends of the remaining arms, AuBe p-type ohmic contacts were formed.

To process the bottom gate the device underwent the epoxy bond and stop-etch (EBASE) technique\cite{Weckwerth1996}.  This  technique entails epoxying the sample to a host substrate topside down and removing material (the original substrate and etch-stop layer) down to the growth superlattice by lapping and etching.  Once EBASE finished, SiN was deposited over the entire mesa surface by plasma enhanced chemical vapor deposition (PECVD) and vias were formed through the epi-material to contact the ohmic pads, now adjacent to the host substrate material.

The back gate metal, TiAu, was then deposited over-top the SiN.  The back gate covers the central Hall bar region, the five mesa arms attached to p-type ohmic contacts and a small area of these contacts.  In this so-called \emph{overlap} configuration, holes are pulled by the backgate into the bottom well from the p-type ohmic contacts\cite{Willet1996}; this is analogous to the n-type ohmic contacts' role described above.  While both quantum wells are physically in contact with the n-type and p-type metal contacts, the gates and mesa configuration is such that only one type carrier is induced in each well.

\section{Experiment}
\label{Section2b}

A schematic of the energy band diagram of the uEHBL during typical operation is given in Fig. \ref{fig0}b.  To simultaneously establish a 2DEG and 2DHG in the uEHBL devices three different, negative bias voltages, top-gate bias $V_{TG}$, bottom-gate bias $V_{BG}$ and interlayer bias $V_{IL}$, are necessarily used.  All these voltages are referenced to ground and at least one 2DHG contact always remains grounded during operation.  The 2DEG is held at $V_{IL}$, which accounts for the difference in the electron and hole Fermi levels and determines E$_{IL}$.  The $V_{IL}$ ends up being less ($\sim$ 35 meV) than the GaAs bandgap energy ($\sim$ 1.51 eV), due to the other field sources, $V_{TG}$, $V_{BG}$ and the carriers in each well.  Ideally, $n$ and $p$ are controlled only by their nearest gate, the top and bottom gates, respectively, due to screening.  However, the system is over-determined (two densities and three voltages) so the same densities can be achieved at different gate voltage settings.  With the 2DEG held at $V_{IL}$ with respect to ground, all the circuitry connected to it must also be held at $V_{IL}$, necessitating the use of an isolation transformer to break the ground of the signal source.

The $n$ and $p$ in the uEHBL were set by adjusting $V_{TG}$, $V_{BG}$ and $V_{IL}$ and measured using low-field Hall measurements.  To characterize transport, the resistivity $\rho$ was measured in each layer as a function of $n$ and $p$, E$_{IL}$ and temperature $T$.  The mobility in each layer was calculated from the resistivity and density according to $\mu_{p} = 1/pe\rho_{p}$ and $\mu_{n} = 1/ne\rho_{n}$.  Resistivity and Hall measurements were made by standard van der Pauw methods using low-frequency, lock-in technique with separate 20 nA excitation currents in both layers.  Coulomb drag measurements were used to estimate the change in $d$ as $V_{IL}$ changed.  For these measurements a 10 nA current was driven in the 2DEG, while the induced voltage in the 2DHG was measured with a high-impedance detection circuit.  The constant temperature measurements were all taken at $T = 0.3$ K in a He$^{3}$ refridgerator.

\section{Results \& Discussion}
\label{Section3}

\begin{figure}
\begin{center}
\includegraphics[width = 9.5cm]{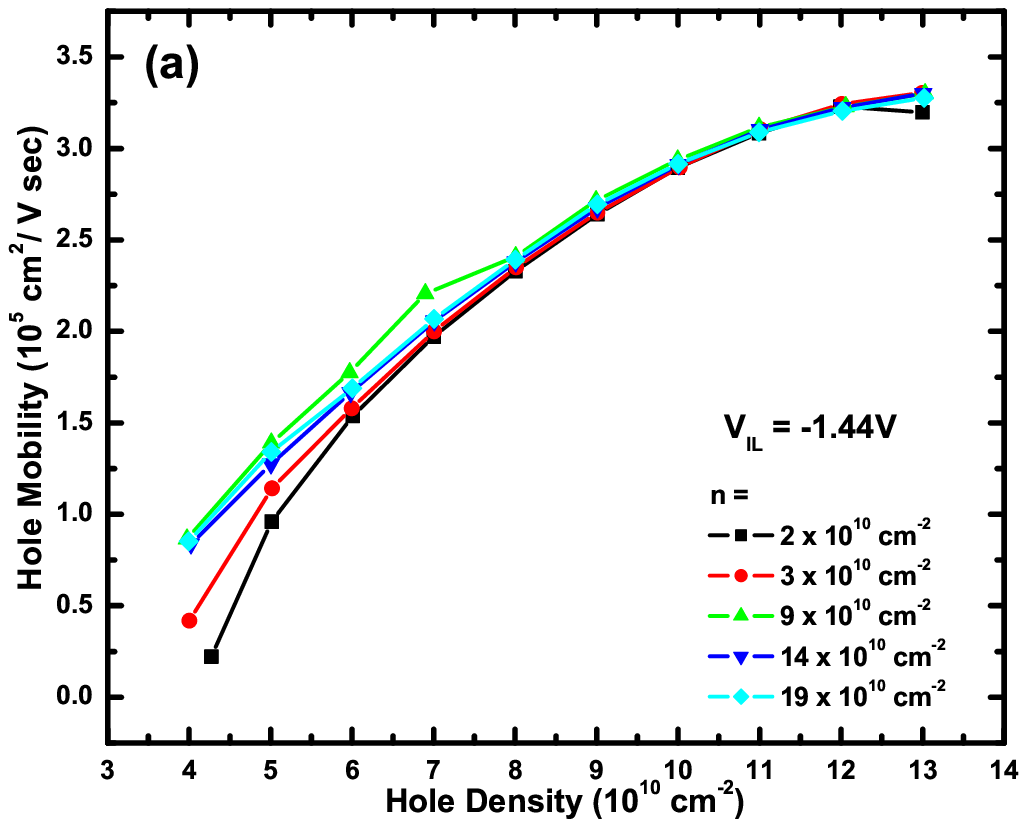}
\includegraphics[width = 9.5cm]{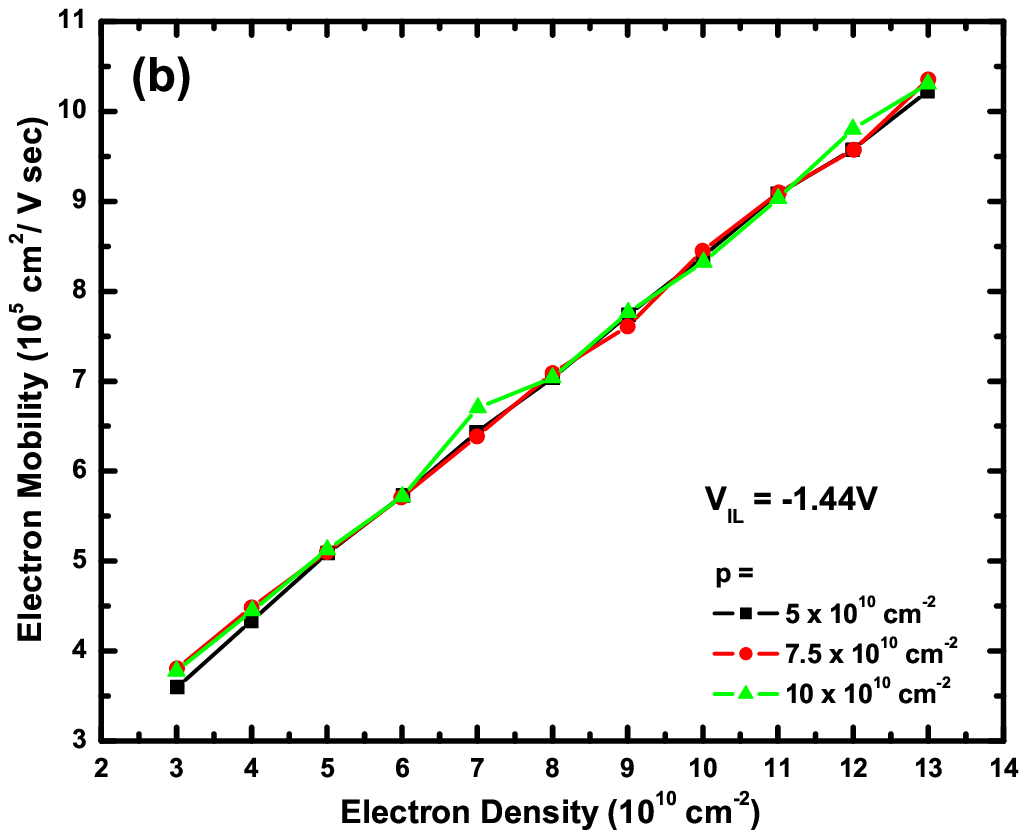}
\end{center}
\caption{\label{fig1} Mobilities (a) $\mu_{p}$ and (b) $\mu_{n}$ as a function of $p$ and $n$, respectively, and the adjacent well density at $T=0.3$ K and $V_{IL} = -1.44$ V.}
\end{figure}

To establish some basis for comparison, the mobility of each layer is plotted in Fig. \ref{fig1} as a function of its density, $n$ or $p$, at different densities in the adjacent layer with $V_{IL} = -1.44$ V.  The resulting interlayer electric field E$_{IL}$ ($\approx 96$ kV/cm) was below the expected breakdown-field in Al$_{.9}$Ga$_{.1}$As, equal to $500$ kV/cm for low temperature.\cite{Ma2002} In the plots of Fig. \ref{fig1}a an increase in hole mobility $\mu_{p}$ with increasing $n$ is visible with a weakening dependence as $p$ increases. The plots of $\mu_{p}$  range from $.25$ to $3.25\times10^{5}$ cm$^{2}$/V s for $p=4.0$ to $13.0\times10^{10}$ cm$^{-2}$.  For $m_{p}^{*} = .45~m_{e}$ this led to a hole scattering time $\tau_{p}$ ranging between $\approx6.4$ and $83.2$ ps. The $\mu_{p}$ did not have an ideal power law $\mu \propto p^{\alpha_{p}}$ dependence on $p$ across this range.  However, the conductivity $\sigma_{p}$ used to calculate $\mu_{p}$ was fit (not shown) to a percolation model ($\sigma_{p} \propto (p-p_{c})^{\gamma_{p}}$). The fit parameter $\gamma_{p}$ increased from $\approx$ $.98$ to $1.12$ and the critical hole density $p_{c}$ decreased from $4.1$ to $3.0\times10^{10}$ cm$^{-2}$ as $n$ increased from $2.0$ to $19.0\times10^{10}$ cm$^{-2}$.  This percolation model was previously used to describe the 2D metal-insulator transition (MIT) in undoped heterostructures.\cite{DasSarma2005}  This range of $p_{c}$ values are below all the $p$ values used in this study so a 2D MIT was not expected in the 2DHG.


In contrast, in the plots in Fig. \ref{fig1}b the electron mobility $\mu_{n}$ shows no dependence on $p$, the density in its adjacent well.  The $\mu_{n}$ ranges from $3.5$ to $10.25\times10^{5}$cm$^{2}$/V s for $n=3.0$ to $13.0\times10^{10}$ cm$^{-2}$. For $m_{n} = .067~m_{e}$, the $\mu_{n}$ led to $\tau_{n}$ between $\approx13.0$ and $39.0$ ps.  In comparison with $\mu_{p}$, $\mu_{n}$ was $\propto n^{\alpha_{n}}$ with $\alpha_{n}=1.02$; this fit used a zero-density offset $\mu_{0} = 1.9\times10^{5}$cm$^{2}$/V s.  Fitting the conductivity $\sigma_{n}$ (not shown) to a similar percolation transport model as above, a fit parameter $\gamma_{n} \approx$ 1.8 was found for all $p$.  The $\sigma_{n}$ fitting, however, indicated the 2DEG will never enter an insulating regime ($n \ll n_{c}$).



\begin{figure}
\begin{center}
\includegraphics[width = 9.5cm]{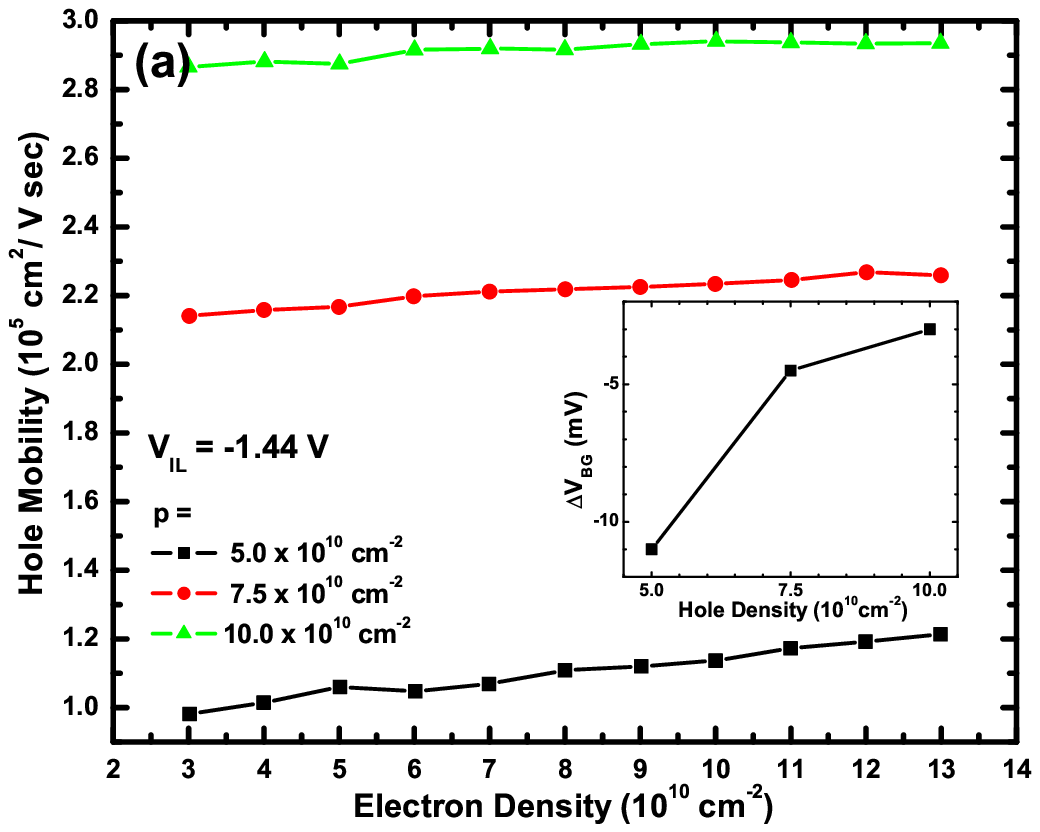}
\includegraphics[width = 9.5cm]{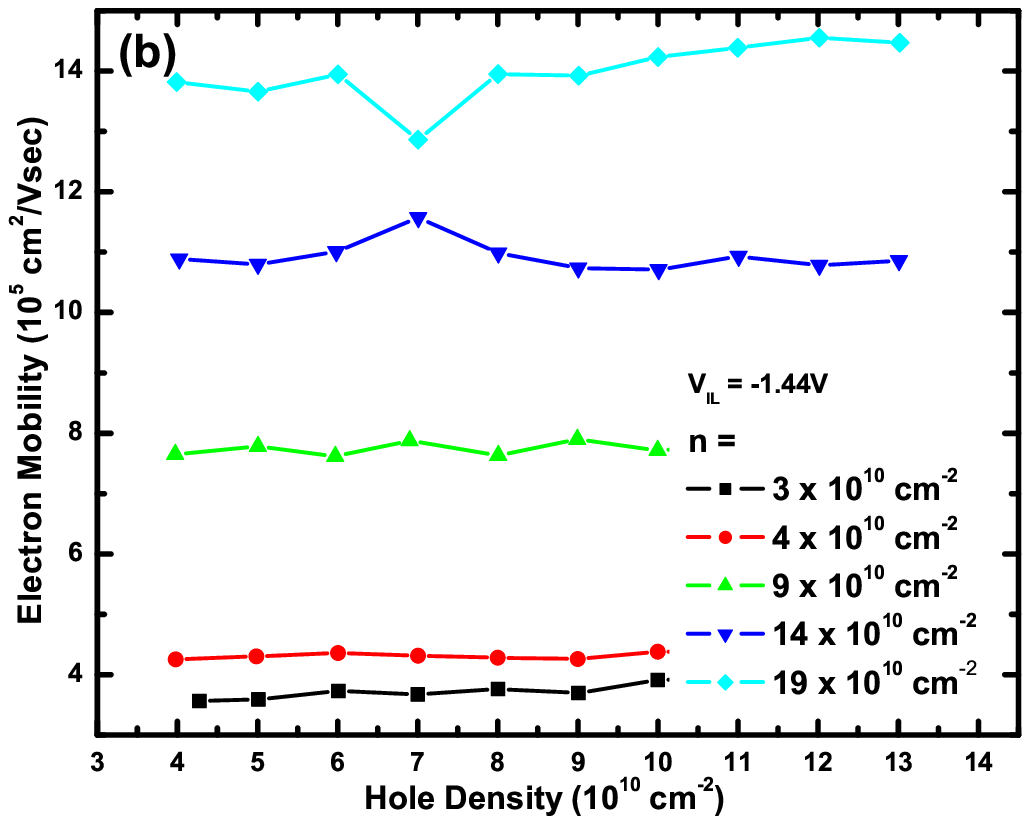}
\end{center}
\caption{\label{fig2} Mobilities (a) $\mu_{p}$ and (b) $\mu_{n}$ as a function of adjacent well carrier density, $n$ and $p$, respectively, at $T=.3$ K and $V_{IL} = -1.44$ V.  The inset plot shows $\Delta V_{BG}$ as a function of $p$ for the data in (a).}
\end{figure}

In Fig. \ref{fig2} the mobility layer interdependence is more closely investigated.  Fig. \ref{fig2}a shows a monotonic relationship existed between $\mu_{p}$ and $n$, the density in the adjacent well.  Furthermore, this dependence apparently becomes weaker, illustrated by the visible decrease in slope between the datasets, as $p$ increased from $5.0$ to $10.0\times10^{10}$ cm$^{-2}$.  This decrease in slope correctly corresponds to the data in Fig. \ref{fig1}a, where the spread between plots was much larger for smaller $p$.  The inset plot of Fig. \ref{fig2}a shows the change in backgate voltage $\Delta V_{BG}$ required to maintain a constant $p$ while $n$ was increased from $3.0$ to $13.0\times 10^{10}$ cm$^{-2}$ using $V_{TG}$.  The traces in Fig. \ref{fig2}b show that $\mu_{n}$ was roughly independent of $p$, confirming what was apparent in Fig. \ref{fig1}b.  Any non-linearity in this data was attributed to noise in the measurements.

A forthcoming, full calculation for mobility-density data on this structure was qualitatively consistent with transport in each layer being dominated by a uniform background (channel) impurity density\cite{DasSarma2007}, which the results above are also consistent with.  The increasing $\mu_{p}$ in Fig. \ref{fig2}a would result from increased squeezing of the hole wavefunction.  Squeezing is known to reduce channel impurity scattering in undoped heterostructures in this density range ($<10^{11}$ cm$^{-2}$).\cite{Pettersen1996}  In the uEHBL, the hole wavefunction squeezing increases with increasing $n$ because $V_{BG}$, which tilts the 2DHG confinement potential, was simultaneously decreased (made more negative) to maintain a constant $p$, as shown in the inset of Fig. \ref{fig2}a.  Furthermore, the decrease in steepness of the slopes of each dataset in Fig. \ref{fig2}a as $p$ was increased is also consistent since the change in $\mu_{p}$ is proportional to $-\Delta V_{BG}$.




If channel impurity scattering also limited the 2DEG mobility then the results in Fig. \ref{fig2}b, that $\mu_{n}$ is roughly independent of $p$, imply squeezing of the electron wavefunction was weaker than squeezing of the hole wavefunction as density in the respective, adjacent well increased.  A direct comparison of the changes in the top and bottom gate voltages is to a large degree voided by the device's asymmetry with regard to the cladding layer widths and relatively different gate leakages, which was a function of the type of gate and contact combinations for either 2D system (see Fig. \ref{fig0}a).  However, the mobility data in Fig. \ref{fig1} still points to channel impurity scattering limiting the 2DEG transport.  From the power-law dependence of $\mu_{n}$ on $n$, the $\alpha_{n}=1.02$ is clearly below the $\alpha=\frac{3}{2}$ expected if remote impurity scattering is limiting the 2DEG.\cite{Pettersen1996}  Furthermore, according to the data in Fig. \ref{fig1} the 2DHG scattering time $\tau_{p}$ grows larger than the 2DEG scattering time $\tau_{n}$ above $n = p \geq 5\times10^{10}$cm$^{-2}$.  For example,  a mobility ratio $\mu_{n}/\mu_{p} \approx 3.3$ exists at $n=p=7.0\times10^{10}$ cm$^{-2}$. This leads to a scattering time ratio $\tau_{p}/\tau_{n}\approx 2.1$.  If channel impurity scattering limits $\mu_{n}$ than this ratio suggests the 2DEG wavefunction is much wider than the hole wavefunction (see eqns. 4-7 in \cite{Kurobe1993}) since that would increase the differential cross section for scattering.  A wider 2DEG wavefuncion is expected since $m_{n}^{*}<m_{p}^{*}$.

\begin{figure}
\begin{center}
\includegraphics[width = 9.5cm]{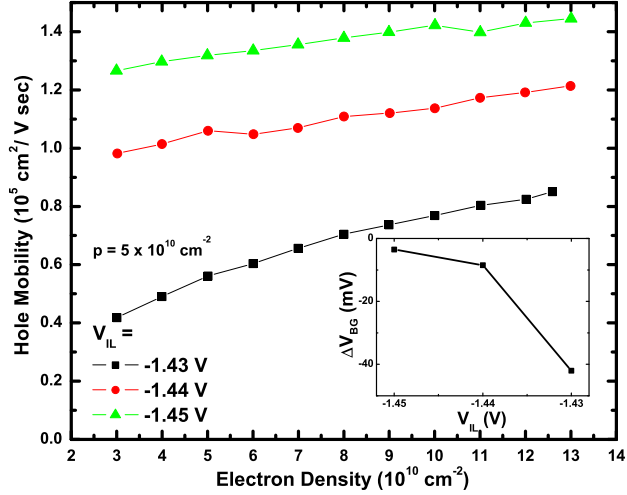}
\end{center}
\caption{\label{fig3} Mobility $\mu_{p}$ as a function of $n$ for $V_{IL}=-1.43$ V to $-1.45$ V and (inset) $\Delta V_{BG}$ as a function of $V_{IL}$ at $T=.3$ K.}
\end{figure}

\begin{figure}
\begin{center}
\includegraphics[width = 9.5cm]{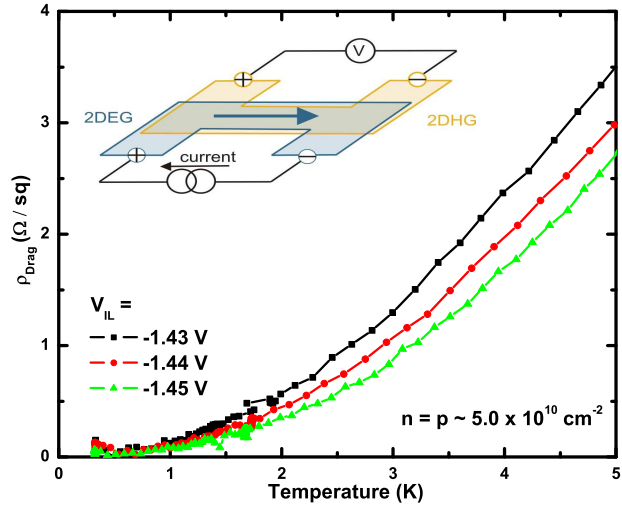}
\end{center}
\caption{\label{fig4} The hole drag resistivity $\rho_{drag}$ as a function of $T$ at matched density $n = p \sim5.0\times10^{10}$ cm$^{_2}$ for various $V_{IL}$ and (inset) schematic of the drag measurement.}
\end{figure}

In Fig. \ref{fig3} the same $\mu_{p}$ data at $p=5\times10^{10}$ cm$^{-2}$ and $V_{IL}=-1.44$ V from Fig. \ref{fig2}a is plotted alongside similar measurements at $V_{IL}=-1.45$ V and $-1.43$ V.  Measurements of $\mu_{n}$ under similar conditions (not shown) were also taken, but showed no discernable dependence on $V_{IL}$.  The inset plot shows the $\Delta V_{BG}$ required to maintain constant $p$ while increasing $n$ from $3.0$ to $12.0 \times 10^{10}$ cm$^{-2}$ at each $V_{IL}$.  With the 2DHG held at ground, making $V_{IL}$, the voltage dropped across the barrier, less negative pulled the 2DEG energy level down, thereby increasing E$_{IL}$ and also ostensibly decreasing interlayer separation $d$ (see Fig. \ref{fig4}).  Based on the prior discussion above, the slopes of the datasets in Fig. \ref{fig3} suggest the largest $\Delta V_{BG}$ would have occurred at $V_{IL}=-1.43$ V since it has the steepest slope and that $\Delta V_{BG}$ increases with decreasing $V_{IL}$.  Both likelihoods are confirmed by the data in the inset plot.

Following the current argument, the increase in $\mu_{p}$ as $V_{IL}$ decreases for fixed $n$ in Fig. \ref{fig3} also implies increased hole wavefunction deformation.  However,  the reverse, an increase in $V_{BG}$ as $V_{IL}$ decreased, was observed.  For example, at $n=3\times10^{10}$ cm$^{-2}$ a $V_{BG} = -1.647$ V resulted for $V_{IL} = -1.43$ V, which was less than $V_{BG} = -1.5735$ V at $V_{IL}=-1.45$ V.  This result suggests the effect of $V_{IL}$ on squeezing the hole wavefunction overcomes the reverse effect from an increasing $V_{BG}$ to enhance $\mu_{p}$ and that narrowing the analysis to only $V_{BG}$'s influences is insufficient to fully account for the device's behavior.



The 2DHG Coulomb drag $\rho_{drag}$ measurements, shown in Fig. \ref{fig4}, were taken as function of $T$ at various $V_{IL}$ for matched density $n=p \sim 5.0\times10^{10}$ cm$^{-2}$.  At $T=0.3$ K a $\rho_{drag}\approx .1~\Omega/$sq was equivalently measured for each $V_{IL}$.  Using $\rho_{drag}=m^{*}_{p}/e^{2}p \tau_{h \rightarrow e}$ the time it takes for a hole to transfer its momentum to an electron is $\tau_{h \rightarrow  e}\approx 313$ ns.\cite{Gramila1991}  This was much longer than the hole scattering time $\tau_{p}$, which varies from $\sim 12.7$ to $33.2$ ps as $V_{IL}$ decreases at $n=p=5\times10^{10}$ cm$^{-2}$ in Fig. \ref{fig3}.

The $\rho_{drag}$ results in Fig. \ref{fig4} also provide some indication that decreasing $V_{IL}$ squeezes the hole wavefunction and, thus, how despite the increase in $V_{BG}$ the $\mu_{p}$ still increased with decreasing $V_{IL}$ at fixed $n$ in Fig. \ref{fig3}.  Above $T = 1.5$ K, $\rho_{drag}$ decreased as $V_{IL}$ was decreased, which is expected, based on theory, if $d$ increased.\cite{Gramila1991}  To analyze this result the ratio $(\rho_{drag}(A)/\rho_{drag}(B))^1/4$ was determined, where $A$ and $B$ were the various $V_{IL}$ and $A>B$.  For constant density this ratio is proportional to the ratio of interlayer separation $d(B)/d(A)$ at each $V_{IL}$.  From the ratio calculation, the decrease $\Delta V_{IL} = -10$ mV  led to an increase $\Delta d\approx5\%$.  For a nominal separation $d = 38$ nm this equates to $1.9$ nm.  This increase in $d$ is suggestive of the hole wavefunction moving away from the 2DEG and closer to the edge of the confinement potential causing squeezing to occur.  Thus, a decrease in $V_{IL}$ also suggests an increase in squeezing and a commensurate increase in $\mu_{p}$.

\begin{figure}
\begin{center}
\includegraphics[width = 9.5cm]{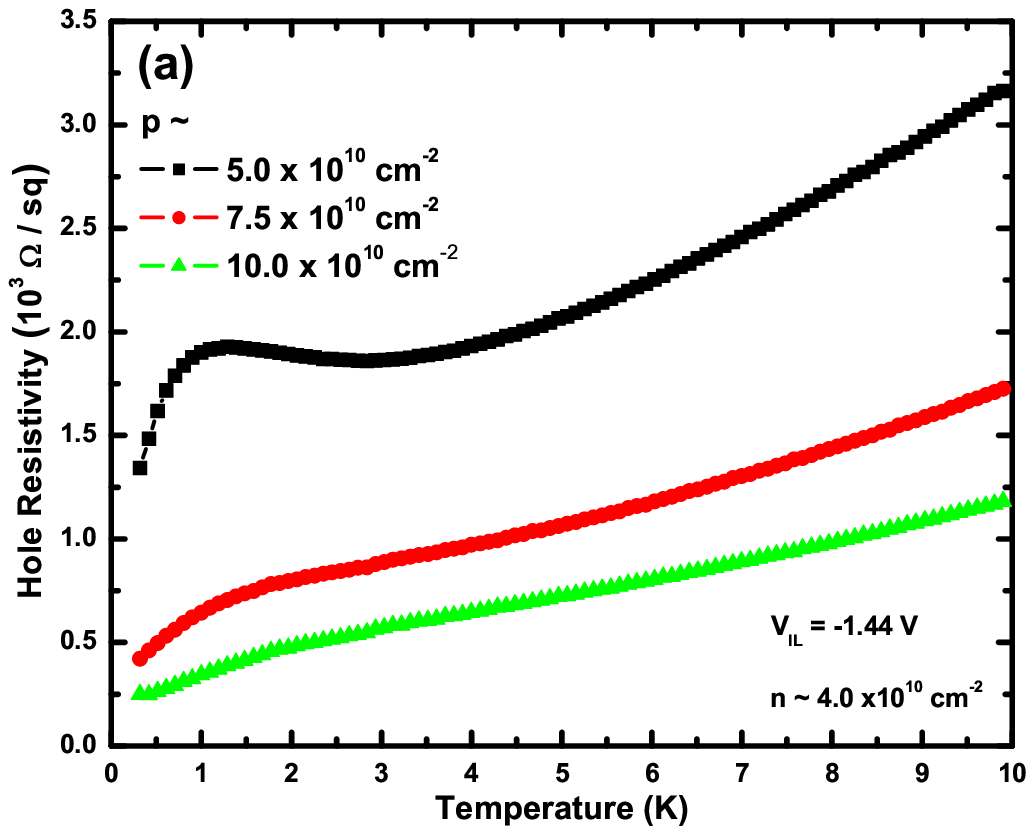}
\includegraphics[width = 9.5cm]{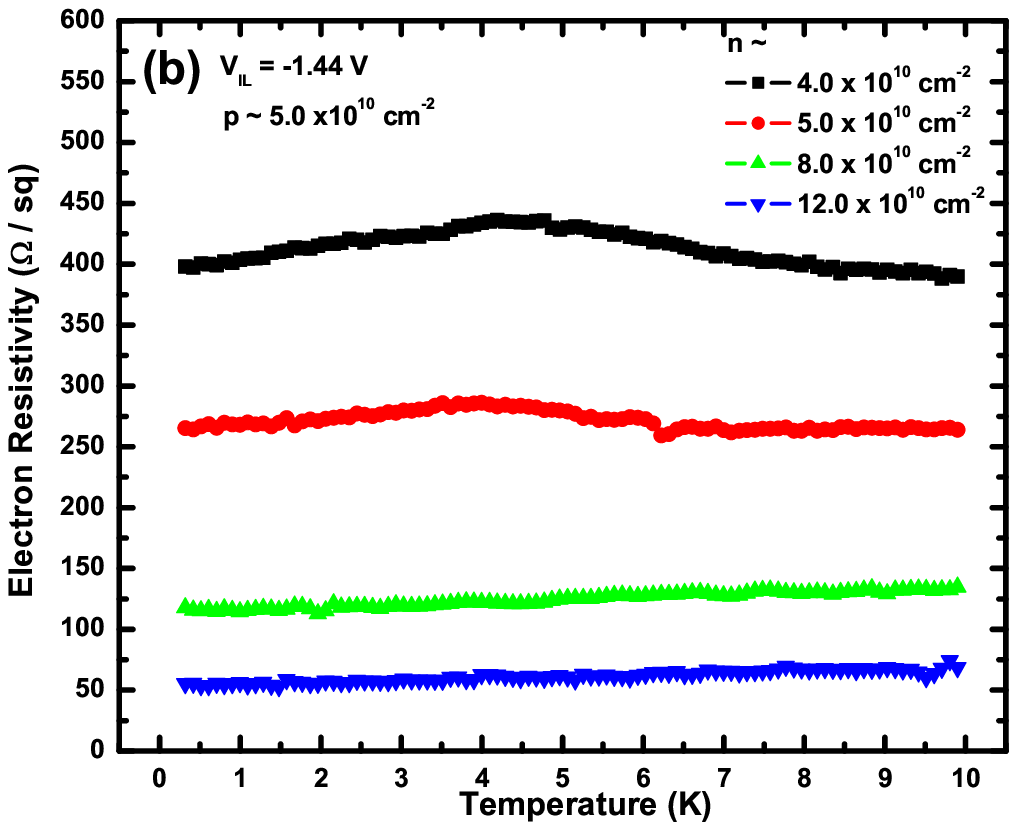}
\end{center}
\caption{\label{fig6} The (a) $\rho_{p}$ and (b) $\rho_{e}$ as a function of $T$ at various $p$ and $n$, respectively, for $V_{IL} = -1.44$ V.}
\end{figure}



Temperature sweeps of $\rho_{p}(T)$ and $\rho_{n}(T)$ as a function of $n$ and $p$ were also taken.  The $\rho_{p}(T)$ results, in Fig. \ref{fig6}a, demonstrate qualitative changes towards an insulator state as $p$ decreased, but not a full transition.  Recall, $p_{c}$ ranging from $\approx 3.0$ to $4.0\times10^{10}$ cm$^{-2}$ was expected from percolation model fittings of $\sigma_{p}$ versus $p$.  The $\rho_{n}(T)$ results in Fig. \ref{fig6}b show qualitatively more metallic behavior and the disappearance of the peak at $T = 4.2$ K as $n$ increased.  Corresponding temperature sweeps of $\rho_{n}(T)$ and $\rho_{p}(T)$ as a function of the adjacent well density, $p$ and $n$, respectively, were also done (not shown).  For $T<4$ K, the former reflected the $\mu_{n}$ data at $T=0.3$ K in Fig. \ref{fig2}b.  For $T>4$ K, however, $\rho_{n}(T, p)$ became smaller as $p$ increased.  The $\rho_{p}(T,n)$ displayed no qualitative changes as $n$ increased from $4.0$ to $12.0\times10^{10}$ cm$^{-2}$ aside from a small increase in amplitude, which reflects the increase of $\mu_{p}$ as $n$ increased in Fig. \ref{fig2}a.


\section{Conclusion}
\label{Section4}

The layer interdependence of low temperature transport in a 30 nm barrier uEHBL device was investigated.  An increase in $\mu_{p}$ with increasing $n$ was observed at various $p$, while no change in $\mu_{n}$ with increasing $p$ was apparent at any $n$.  The former appears to be only an indirect effect, however, since $V_{BG}$ was simultaneously decreased to maintain constant $p$ while $n$ was increased.  This led to further tilting of the 2DHG confinement potential, thereby increasing the hole wavefunction squeezing and, commensurately, reducing the dominant background impurity scattering.  Decreasing V$_{IL}$ was observed to increase both $\mu_{p}$ and $d$.  A $\Delta d \approx 5\%$ increase with $\Delta V_{IL}=-10$ mV was determined by $\rho_{drag}(T)$ measurements.  Finally, measurements of $\rho_{p}(T)$ and $\rho_{n}(T)$ showed qualitative hints of a 2DHG transition to an insulator state that disappeared as $p$ increased and of the 2DEG becoming more metallic as $n$ increased, respectively.

\begin{acknowledgments}
\label{Section5}
The author's acknowledge Denise Tibbets for her technical assistance.  The author's also wishes to thank S. Das Sarma and E. H. Hwang for their helpful discussions of the related theory.  This work has been supported by the Division of Materials Sciences and Engineering, Office of Basic Energy Sciences, U.S. Department of Energy. Sandia is a multiprogram laboratory operated by Sandia Corporation, a Lockheed Martin Company, for the United States Department of Energy under Contract No. DE-AC04-94AL85000.
\end{acknowledgments}

\section*{References}

\bibliography{EHBL_Bib}

\end{document}